\documentstyle[12pt,a4]{article}
\newcommand\lsim{\mathrel{\rlap{\lower4pt\hbox{\hskip1pt$\sim$}}
    \raise1pt\hbox{$<$}}}
\newcommand\gsim{\mathrel{\rlap{\lower4pt\hbox{\hskip1pt$\sim$}}
    \raise1pt\hbox{$>$}}}

\def\k{{\bf k}}

\catcode`\@=11
\def\marginnote#1{}
\def\ifmath#1{\relax\ifmmode #1\else $#1$\fi}

\def\stop{\,\widetilde{t}}

\def\bold#1{\setbox0=\hbox{$#1$}%
     \kern-.025em\copy0\kern-\wd0
     \kern.05em\copy0\kern-\wd0
     \kern-.025em\raise.0433em\box0 }

\def\GENITEM#1;#2{\par\vskip6pt \hangafter=0 \hangindent=#1
   \Textindent{$ #2$ }\ignorespaces}

\newcount\hour
\newcount\minute
\newtoks\amorpm
\hour=\time\divide\hour by60
\minute=\time{\multiply\hour by60 \global\advance\minute by-
\hour}
\edef\standardtime{{\ifnum\hour<12 \global\amorpm={am}%
    \else\global\amorpm={pm}\advance\hour by-12 \fi
    \ifnum\hour=0 \hour=12 \fi
    \number\hour:\ifnum\minute<100\fi\number\minute\the\amorpm}}
\edef\militarytime{\number\hour:\ifnum\minute<100\fi\number\minute}
\def\draftlabel#1{{\@bsphack\if@filesw {\let\thepage\relax
  \xdef\@gtempa{\write\@auxout{\string
    \newlabel{#1}{{\@currentlabel}{\thepage}}}}}\@gtempa
    \if@nobreak \ifvmode\nobreak\fi\fi\fi\@esphack}
     \gdef\@eqnlabel{#1}}
\def\@eqnlabel{}
\def\@vacuum{}
\def\draftmarginnote#1{\marginpar{\raggedright\scriptsize\tt#1}}
\def\draft{\oddsidemargin -.5truein
        \def\@oddfoot{\sl preliminary draft \hfil
        \rm\thepage\hfil\sl\today\quad\militarytime}
        \let\@evenfoot\@oddfoot \overfullrule 3pt
        \let\label=\draftlabel
        \let\marginnote=\draftmarginnote

\def\@eqnnum{(\theequation)\rlap{\kern\marginparsep\tt\@eqnlabel}%
\global\let\@eqnlabel\@vacuum}  }
\def\preprint{\twocolumn\sloppy\flushbottom\parindent 1em
        \leftmargini 2em\leftmarginv .5em\leftmarginvi .5em
        \oddsidemargin -.5in    \evensidemargin -.5in
        \columnsep 15mm \footheight 0pt
        \textwidth 250mmin      \topmargin  -.4in
        \headheight 12pt \topskip .4in
        \textheight 175mm
        \footskip 0pt

\def\@oddhead{\thepage\hfil\addtocounter{page}{1}\thepage}
        \let\@evenhead\@oddhead \def\@oddfoot{} \def\@evenfoot{}
}
\def\titlepage{\@restonecolfalse\if@twocolumn\@restonecoltrue\o
necolumn
     \else \newpage \fi \thispagestyle{empty}\c@page\z@
        \def\thefootnote{\fnsymbol{footnote}} }
\def\endtitlepage{\if@restonecol\twocolumn \else  \fi
        \def\thefootnote{\arabic{footnote}}
        \setcounter{footnote}{0}}  
\catcode`@=12
\relax
\def\be{\begin{equation}}
\def\ee{\end{equation}}
\def\bea{\begin{eqnarray}}
\def\eea{\end{eqnarray}}

\def\simgt{\stackrel{>}{{}_\sim}}

\def\NPB#1#2#3{{\it Nucl.~Phys.} {\bf{B#1}} (19#2) #3}
\def\PLB#1#2#3{{\it Phys.~Lett.} {\bf{B#1}} (19#2) #3}
\def\PRD#1#2#3{{\it Phys.~Rev.} {\bf{D#1}} (19#2) #3}
\def\PRL#1#2#3{{\it Phys.~Rev.~Lett.} {\bf{#1}} (19#2) #3}
\def\ZPC#1#2#3{{\it Z.~Phys.} {\bf C#1} (19#2) #3}

\def\AP#1#2#3{{\it Ann.~Phys.} {\bf#1} (19#2) #3}
\def\RMP#1#2#3{{\it Rev.~Mod.~Phys.} {\bf#1} (19#2) #3}

\def\mst11{m_{\;\widetilde{t}_{1}}}

\def\mst22{m_{\;\widetilde{t}_{2}}}
\def\mst12{m_{\;\widetilde{t}_{1,2}}}

\def\msb11{m_{\;\widetilde{b}_{1}}}
\def\msb22{m_{\;\widetilde{b}_{2}}}
\def\msb12{m_{\;\widetilde{b}_{1,2}}}

\def\mwidetilde2{\widetilde{m}^{2}}

\relax

\begin{document}
\topmargin-2.5cm
%
\begin{titlepage}
\begin{flushright}
CERN-TH/97-348\\
OUTP-97-71-P\\
\end{flushright}
\vskip 0.3in
\begin{center}
{\Large\bf  Supersymmetric Electroweak Baryogenesis, }
\vskip 0.1in
{\Large\bf Nonequilibrium Field Theory and}
\vskip 0.1in
{\Large\bf Quantum Boltzmann Equations  }
\vskip .5in

{\large\bf A. Riotto\footnote{On leave of absence from Department of Physics, Theoretical  Physics, University of Oxford, U.K.}    }
\vskip.35in
 CERN, TH Division\\
CH-1211 Geneva 23, Switzerland 
\end{center}
\vskip 0.5cm
\begin{center}
{\bf Abstract}
\end{center}
\begin{quote}
The closed time-path (CTP) formalism is a powerful Green's function formulation to describe nonequilibrium phenomena in field theory and it leads to a complete nonequilibrium quantum kinetic theory. In this paper we make use of the CTP formalism to write down a set of quantum Boltzmann equations describing the local number  density asymmetries of the particles involved in supersymmetric electroweak baryogenesis. These diffusion equations  automatically and self-consistently incorporate  the CP-violating sources which fuel baryogenesis when transport properties allow the CP-violating charges to diffuse in front of the bubble wall separating the broken from the unbroken phase at the electroweak phase transition. This is a significant improvement with respect to  recent approaches where the CP-violating sources are  inserted by hand into the diffusion equations. Furthermore, the CP-violating sources and the particle number changing interactions manifest ``memory'' effects which are typical of the quantum transport theory and are not present in the classical approach. The slowdown of the relaxation processes  may  keep the system out of equilibrium for longer times and therefore enhance the final baryon asymmetry. We also stress that the classical approximation is not adequate to describe the quantum interference nature of  CP-violation and that a  quantum approach should be adopted to compute the sources since they are most easily built up by the transmission of low momentum particles.

\end{quote}
\vskip 0.5cm
\begin{flushleft}
December 1997 \\
\end{flushleft}

\end{titlepage}
\setcounter{footnote}{0}
\setcounter{page}{0}
\newpage
\baselineskip=20pt

\begin{flushleft}
{\bf  1. Introduction and summary}
\end{flushleft}

Because of  the presence of  unsuppressed
baryon number violating processes at high temperatures, 
the Standard Model (SM) of weak interactions fulfills all the requirements for  a successful   
generation of the baryon number at the electroweak scale \cite{reviews}. The baryon number violating processes    also impose severe constraints on  
models where the baryon asymmetry is created at energy scales much higher than the electroweak scale \cite{anomaly}. Unfortunately, the electroweak phase transition  is too weak  in the SM \cite{transition}. This   means that the baryon asymmetry  
generated during the transition would be subsequently erased by unsuppressed  
sphaleron transitions in the broken phase. 
 The most promising and  
well-motivated framework for electroweak baryogenesis beyond the SM  seems to be supersymmetry (SUSY).  Electroweak  
baryogenesis in the framework of the Minimal Supersymmetric Standard Model  
(MSSM) has  attracted much attention in the past years with 
particular emphasis on the strength of the phase transition ~\cite{early1,early2,early3} and  
the mechanism of baryon number generation \cite{nelson,noi,higgs,ck}.

Recent  analytical \cite{r1,r2} and  lattice  
computations  \cite{r3} have  revealed   that the phase transition can be sufficiently strongly  
first order if  the
ratio of the vacuum expectation values of the two neutral Higgses $\tan\beta$  
is smaller than $\sim 4$. Moreover, taking into account all the experimental bounds 
as well as those coming from the requirement of avoiding dangerous
 color breaking minima,  the lightest Higgs boson should be  lighter than about  $105$ GeV,
 while the right-handed stop mass might  be close to the present experimental bound and should
 be smaller than, or of order of, the top quark mass \cite{r2}. 

Moreover, the MSSM contains additional sources  
of CP-violation  besides the CKM matrix phase. 
These new phases are essential  for the generation of the baryon number since  large  
CP-violating sources may be  locally induced by the passage of the bubble wall separating the broken from the unbroken phase during the electroweak phase transition.   Baryogenesis is fueled  when transport properties allow the CP-violating  
charges to efficiently diffuse in front of the advancing bubble wall where
anomalous electroweak baryon violating processes are not suppressed.
The new phases   
appear    in the soft supersymmetry breaking  
parameters associated to the stop mixing angle and to  the gaugino and  
neutralino mass matrices; large values of the stop mixing angle
are, however, strongly restricted in order to preserve a
sufficiently strong first order electroweak phase transition. 
Therefore, an acceptable baryon asymmetry from the stop sector
may only be generated through a delicate balance between the values
of the different soft supersymmetry breaking parameters contributing
to the stop mixing parameter, and their associated CP-violating 
phases \cite{noi}. As a result, the contribution to the final baryon asymmetry from the stop sector turns out to be negligible.   On the other hand, charginos and neutralinos may be responsible for the observed baryon asymmetry if   the phase of the parameter $\mu$ is    large enough \cite{noi,ck}. Yet, 
this is true within the MSSM. If the strength of the
 electroweak phase transition is enhanced by the presence of some
new degrees of freedom beyond the ones contained in the MSSM, {\it
e.g.} some extra standard model gauge singlets,
 light stops (predominantly the 
right-handed ones) and charginos/neutralinos are expected to 
give quantitatively the
same contribution to the final baryon asymmetry.  

The baryon asymmetry has been usually computed using the following  separate steps \cite{nelson,noi,riotto}:

{\it 1)}  Look for 
 those  charges which are
approximately conserved in the symmetric phase, so that they
can efficiently diffuse in front of the bubble where baryon number
violation is fast, and non-orthogonal to baryon number,
so that the generation of a non-zero baryon charge is energetically
favoured.
Charges with these characteristics in the MSSM 
are the axial stop charge and the Higgsino charge,
which may be produced from the interactions of squarks and
charginos
and/or neutralinos with the bubble wall,
provided a source of CP-violation is present in these sectors.

{\it 2)}  Compute the CP-violating currents  of the plasma locally induced by the passage of the bubble wall. The methods present in the literature properly incorporate the decoherence effects which may have a crucial impact on the generation of the CP-violating observable.  

{\it 3)} Write and solve a set of coupled differential diffusion equations for the local particle densities, including the CP-violating source terms derived from the computation of the current at  step {\it 2)}  and the particle number changing reactions. The solution to these equations gives a net baryon number which is produced in the symmetric phase and then transmitted into the interior of the bubbles of broken phase, where it is not wiped out if the first transition is strong enough. 

It is important to notice that the CP-violating sources are inserted into the diffusion equations by hand only after the CP-violating currents have been defined and computed. 
This procedure is  certainly appropriate to describe the damping effects on the CP-violating observables 
originated by the
plasma interactions, but
does not  incorporate any relaxation time scale
arising when diffusion and particle changing interactions are
included (even though this approximation might be good if the diffusion time scales are larger than the  damping time scales) and is theoretically not consistent. Furthermore, since a certain degree of  arbitrariness is present in the way the CP-violating sources may be defined, different  CP-violating sources have been adopted  for the stop and the Higgsino sectors in the literature \cite{nelson,noi}. This is certainly not an academic question since different sources may lead to   different numerical results for the final baryon asymmetry, especially if  the sources are  expressed in terms of   a different number of derivatives of the Higgs bubble wall profile and, therefore, in terms of  different powers of the bubble wall velocity $v_\omega$ and bubble wall width $L_\omega$. 

It is indisputable  that one might    be able to rigously derive  a set of  transport (diffusion) equations {\it already} incorporating the CP-violating sources in a self-consistent way only by means of a more complete treatment of the problem.  The goal of this paper is to show that    nonequilibrium Quantum Field Theory  provides us with   the necessary    tools  to  write down a set of quantum Boltzmann equations (QBE's) describing the local particle densities and automatically incorporating the CP-violating sources.    The ordinary quantum field theory at finite temperature is not useful to
study the dynamics of  particle densities. This is because we
need their temporal evolution with definite initial conditions and not
simply the transition amplitude of particle reactions with fixed initial
and final conditions. The most appropriate extension of the field theory
to deal with these issues it to generalize the time contour of
integration to a closed time-path (CTP).  The CTP formalism is a powerful Green's function
formulation for describing nonequilibrium phenomena in field theory, it leads to 
 a complete
nonequilibrium quantum kinetic theory approach and  it will guide us towards  the   rigorous   computation of  the CP-violating sources for the stop and the Higgsino numbers.   This will also eliminate the level  of  arbitrariness  the previous treatments are suffering from. 

There exist other good reasons why one should call for  the  nonequilibrium quantum kinetic theory. 
The fact that CP-violating sources  are     
most  easily built up  by the transmission of low momentum particles over a distance $L_\omega$ \cite{nelson,noi}  is an indication that  particles with masses smaller than or of the order of the temperature $T$ are  relevant in the process of quantum interference  leading to CP-violating sources  in the bubble wall.  Basically,  the sources are  dominated by particles with long wavelengths in direction perpendicular to the wall. 
This tells us  that the classical approximation is not adequate to
describe the quantum interference nature of $CP$-violation and a quantum approach must be adopted to compute the sources.  For low momentum particles, 
the validity of the {\it classical} Boltzmann equation  starts to break down and the ultimate answer can
be provided only by a complete nonequilibrium quantum field theory
approach. Kinetic theory and classical Boltzmann equations have been used to
describe the dynamics of particles treated as classical with a defined
position, energy and momentum.  This requires that, in
particular, the mean free path must be large compared to the Compton
wavelength of the underlying particle in order for the classical picture
to be valid, which is not guaranteed for
particles with a small momentum perpendicular to the
wall. Distribution functions obeying the quantum Boltzmann equations are the  only  correct
functions to describe particles in an interacting, many-particle
environment.  Furthermore,  we will show that the CP-violating sources and the particle number changing interactions  built up from the CTP formalism are characterized by  ``memory'' effects which are typical of  the  quantum transport theory \cite{dan,henning}. In  the classical kinetic theory the ``scattering term'' does not include any integral over the past history of the system. This   is equivalent to assume that any collision in the plasma  does not depend upon the previous ones. On the contrary,   
quantum distributions posses strong memory effects which are relevant for the computation of the final baryon asymmetry since they lead to a slowdown of   thermalization times and therefore to longer stages of nonequilibrium. 

The paper is organized as follows. In  section 2  we give a brief description of the basic features of the nonequilibrium quantum field theory and the CTP formalism. In sections 3  and 4  we compute the quantum transport equations for local particle asymmetries in the bosonic and fermion case, respectively. Sections 5  and  6  contain the explicit 
computation of the CP-violating sources for the right-handed stop  and the Higgsino numbers.  We conclude with an outlook of our findings and comments about  their implications in section 7.

\begin{flushleft}
{\bf  2. Some basics of  non-equilibrium quantum field theory}
\end{flushleft}

In this section  we will   briefly present    some of the  basic features of the  nonequilibrium quantum field theory. The interested reader is referred to the excellent review by Chou  et {\it al.} \cite{chou} for a more exhaustive discussion.

The ordinary quantum field theory at finite temperature, which mainly
deals with transition amplitudes in particle reactions, is not useful to
study the dynamics of  particle asymmetries. This is because we
need their temporal evolution with definite initial conditions and not
simply the transition amplitude of particle reactions with fixed initial
and final conditions. 

The most appropriate extension of the field theory
to deal with these issues it to generalize the time contour of
integration to a closed-time path. More precisely, the time integration
contour is deformed to run from $-\infty$ to $+\infty$ and back to
$-\infty$.  

 The CTP formalism (often  dubbed as in-in formalism) is a powerful Green's function
formulation for describing nonequilibrium phenomena in field theory.  It
allows to describe phase-transition phenomena and to obtain a
self-consistent set of quantum Boltzmann equations.
The formalism yields various quantum averages of
operators evaluated in the in-state without specifying the out-state. On the contrary, the ordinary quantum field theory (often dubbed as in-out formalism) yields quantum averages of the operators evaluated  with an in-state at one end and an out-state at the other. 

The partition function in the in-in formalism for a {\it complex} scalar field is defined to be
\begin{eqnarray}
Z\left[ J, J^{\dagger}\right] &=& {\rm Tr}\:\left[ T\left( {\rm exp}\left[i\:\int_C\:\left(J\phi+J^{\dagger}\phi^{\dagger} \right)\right]\right)\rho\right]\nonumber\\
&=& {\rm Tr}\:\left[ T_{+}\left( {\rm exp}\left[ i\:\int\:\left(J_{+}\phi_{+}+J^{\dagger}_{+}\phi^{\dagger}_{+} \right)\right]\right)\right.
\nonumber\\
&\times&\left.  T_{-}\left( {\rm exp}\left[ -i\:\int\:\left(J_{-}\phi_{-}+J^{\dagger}_{-}\phi^{\dagger}_{-} \right)\right]\right) \rho\right],
\end{eqnarray}
where the suffic $C$ in the integral denotes that the time integration contour runs from minus infinity to plus infinity and then back to minus infinity again. The symbol $\rho$ represents the initial density matrix and the fields are in the Heisenberg picture  and  defined on this closed time contour. 

As with the Euclidean time formulation, scalar (fermionic) fields $\phi$ are
still periodic (anti-periodic) in time, but with
$\phi(t,\vec{x})=\phi(t-i\beta,\vec{x})$, $\beta=1/T$.
The temperature appears   due to boundary
condition, but now time is explicitly present in the integration
contour.

For non-equilibrium phenomena and as a consequence of the time contour, we must now identify field
variables with arguments on the positive or negative directional
branches of the time path. This doubling of field variables leads to
six  different real-time propagators on the contour \cite{chou}.  It is possible to employ fewer  than six since they are not independent, but using six simplifies the notation. 
For a generic bosonic charged  scalar field $\phi$ they are defined as 
\begin{eqnarray}
\label{def1}
G_{\phi}^{>}\left(x, y\right)&=&-i\langle
\phi(x)\phi^\dagger (y)\rangle,\nonumber\\
G_{\phi}^{<}\left(x,y\right)&=&-i\langle
\phi^\dagger (y)\phi(x)\rangle,\nonumber\\
G^t _{\phi}(x,y)&=& \theta(x,y) G_{\phi}^{>}(x,y)+\theta(y,x) G_{\phi}^{<}(x,y),\nonumber\\
G^{\bar{t}}_{\phi} (x,y)&=& \theta(y,x) G_{\phi}^{>}(x,y)+\theta(x,y) G_{\phi}^{<}(x,y), \nonumber\\
G_{\phi}^r(x,y)&=&G_{\phi}^t-G_{\phi}^{<}=G_{\phi}^{>}-G^{\bar{t}}_{\phi}, \:\:\:\: G_{\phi}^a(x,y)=G^t_{\phi}-G^{>}_{\phi}=G_{\phi}^{<}-G^{\bar{t}}_{\phi},
\end{eqnarray}
where the last two Green functions are the retarded and advanced Green functions respectively and $\theta(x,y)=\theta(t_x-t_y)$ is the step function.  For a generic fermion field $\psi$ the six different propagators are analogously defined as
\begin{eqnarray}
\label{def2}
G^{>}_{\psi}\left(x, y\right)&=&-i\langle
\psi(x)\bar{\psi} (y)\rangle,\nonumber\\
G^{<}_{\psi}\left(x,y\right)&=&+i\langle
\bar{\psi}(y)\psi(x)\rangle,\nonumber\\
G^{t}_{\psi} (x,y)&=& \theta(x,y) G^{>}_{\psi}(x,y)+\theta(y,x) G^{<}_{\psi}(x,y),\nonumber\\
G^{\bar{t}}_{\psi} (x,y)&=& \theta(y,x) G^{>}_{\psi}(x,y)+\theta(x,y) G^{<}_{\psi}(x,y),\nonumber\\
G^r_{\psi}(x,y)&=&G^{t}_{\psi}-G^{<}_{\psi}=G^{>}_{\psi}-G^{\bar{t}}_{\psi}, \:\:\:\: G^a_{\psi}(x,y)=G^{t}_{\psi}-G^{>}_{\psi}=G^{<}_{\psi}-G^{\bar{t}}_{\psi}.
\end{eqnarray}
For equilibrium phenomena, the brackets $\langle \cdots\rangle$ imply a thermodynamic average over all the possible states of the system. For homogeneous systems in equilibrium, the Green functions
depend only upon the difference of their arguments $(x,y)=(x-y)$, and there is no dependence upon $(x+y)$. For systems out of equilibrium, the definitions (\ref{def1}) and (\ref{def2}) have a different meaning. The bracket no longer signifies thermodynamic averaging since the concept is now ill-defined. Instead, the bracket means the need to average over all the available states of the system for the non-equilibrium distributions. Furthermore, the arguments of the Green functions $(x,y)$ are {\it not} usually given as the difference $(x-y)$. For example, non-equilibrium could be caused by transients which make the Green functions
depend upon $(t_x,t_y)$ rather than $(t_x-t_y)$. 

For interacting systems whether in equilibrium or not, one must define and calculate self-energy functions. There are six of them: $\Sigma^{t}$, $\Sigma^{\bar{t}}$, $\Sigma^{<}$, $\Sigma^{>}$, 
$\Sigma^r$ and $\Sigma^a$. The same relationships exist among them as for the Green functions in  (\ref{def1}) and (\ref{def2}), such as
\begin{equation}
\Sigma^r=\Sigma^{t}-\Sigma^{<}=\Sigma^{>}-\Sigma^{\bar{t}}, \:\:\:\:\Sigma^a=\Sigma^{t}-\Sigma^{>}=\Sigma^{<}-\Sigma^{\bar{t}}. 
\end{equation}
The self-energies are incorporated into the Green functions through the use of  Dyson's equations. A useful notation may be introduced which expresses four of the six Green functions as the elements of two-by-two matrices \cite{craig}

\begin{equation}
\widetilde{G}=\left(
\begin{array}{cc}
G^{t} & \pm G^{<}\\
G^{>} & - G^{\bar{t}}
\end{array}\right), \:\:\:\:
\widetilde{\Sigma}=\left(
\begin{array}{cc}
\Sigma^{t} & \pm \Sigma^{<}\\
\Sigma^{>} & - \Sigma^{\bar{t}}
\end{array}\right),
\end{equation}
where the upper signs refer to bosonic case and the lower signs to fermionic case. For systems either in equilibrium or non-equilibrium, Dyson's equation is most easily expressed by using the matrix notation
\begin{equation}
\label{d1}
\widetilde{G}(x,y)=\widetilde{G}^0(x,y)+\int\: d^4x_3\:\int d^4x_4\: \widetilde{G}^0(x,x_3)
\widetilde{\Sigma}(x_3,x_4)\widetilde{G}(x_4,y),
\end{equation}
where the superscript ``0'' on the Green functions means to use those for noninteracting system.  This equation is illustrated  in Fig. 1, where the thick solid lines represent the full Green function and the thin solid lines represent the propagators for the noninteracting theory. The expression appears quite formidable; however, some simple expressions may  be obtained for the respective Green functions. It is useful to notice that Dyson's equation can be written in an alternate form, instead of  (\ref{d1}), with $\widetilde{G}^0$ on the right in the interaction terms, see Fig. 2:
\begin{equation}
\label{d2}
\widetilde{G}(x,y)=\widetilde{G}^0(x,y)+\int\: d^4x_3\:\int d^4x_4\: \widetilde{G}(x,x_3)
\widetilde{\Sigma}(x_3,x_4)\widetilde{G}^0(x_4,y).
\end{equation}
Eqs. (\ref{d1}) and (\ref{d2}) are the starting points to derive the quantum Boltzmann equations
describing the temporal evolution of the CP-violating particle density asymmetries.

\begin{flushleft}
{\bf  3. QBE  for particle density asymmetry: the bosonic case}
\end{flushleft}

Kadanoff and Baym \cite{kb}  provided a general method of deriving the QBE's. Here we adopt their  technique and approach  to derive the QBE's for some generic bosonic particle asymmetry. This will allow us to derive in a self-consistent way  the CP-violating sources fueling electroweak baryogenesis in the diffusion equation for the right-handed stop asymmetry. 

Our goal is  to find the QBE   for the following CP-violating current
\begin{equation}
\langle J_\phi^\mu(x) \rangle \equiv i \langle \phi^{\dagger}(x)\stackrel{\leftrightarrow}{\partial}_x^\mu  \phi(x)\rangle\equiv \left[ n_\phi(x), \vec{J}_\phi(x)\right].
\end{equation}
The zero-component of this current $n_\phi$ represents the number density of particles minus the number density of antiparticles and is therefore the quantity which enter the diffusion equations of supersymmetric electroweak baryogenesis. 

Since  the CP-violating current can be expressed in terms of the Green function
$G^{<}_{\phi}(x,y)$ as
\begin{equation}
\label{c1}
\langle J_\phi^\mu(x) \rangle= - \left. \left(\partial_x^\mu - \partial_y^\mu\right) G^{<}_{\phi}(x,y)\right|_{x=y},
\end{equation}
 the problem is reduced to find the QBE for the Green function $G^{<}_{\phi}(x,y)$.  To make contact with the standard  derivation of the QBE \cite{kb}, we may go to a center-of-mass coordinate system
\begin{equation}
\label{dd}
X=(T,\vec{R})=\frac{1}{2}(x+y),\:\:\:\: (t,\vec{r})=x-y.
\end{equation}
Note that $T$ now means the center-of-mass time and not temperature. The notation on the Green function is altered to these center-of-mass coordinates
\begin{equation}
G^{<}_{\phi}(x,y)=G^{<}_{\phi}(t,\vec{r},T,\vec{R})=-i\langle \phi^{\dagger}\left(T-\frac{1}{2}t, \vec{R}-\frac{1}{2}
\vec{r}\right)\phi\left(T+\frac{1}{2}t, \vec{R}+\frac{1}{2}
\vec{r}\right)\rangle.
\end{equation}
The identification $x=y$ in Eq. (\ref{c1}) is therefore equivalent to require $t=\vec{r}=0$. 

Our interest is in finding an equation of motion for the interacting Green function $G^{<}_{\phi}$ when the system in not in equilibrium.  Such an equation can be found from (\ref{d1}) by operating  by
$\left(\stackrel{\rightarrow}{\Box}_x+m^2\right)$ on both sides of the equation. Here $m$ represents the mass term of the field $\phi$. On the right side, this operator acts only on $\widetilde{G}_\phi^0$
\begin{equation}
\label{f1}
\left(\stackrel{\rightarrow}{\Box}_x+m^2\right)\widetilde{G}_\phi(x,y)=\delta^{(4)}(x,y)\widetilde{I}_4+
\int\:d^4 x_3 \widetilde{\Sigma}_\phi(x,x_3)\widetilde{G}_\phi(x_3,y),
\end{equation}
where $I$ is the identity matrix. It is useful to also have an equation of motion for the other variable $y$. This is obtained from (\ref{d2}) by operating  by
$\left(\stackrel{\leftarrow}{\Box}_y+m^2\right)$ on both sides of the equation. We obtain
\begin{equation}
\label{f2}
\widetilde{G}_\phi(x,y)\left(\stackrel{\leftarrow}{\Box}_y+m^2\right)=\delta^{(4)}(x,y)\widetilde{I}_4+
\int\:d^4 x_3 \widetilde{G}_\phi(x,x_3)\widetilde{\Sigma}_\phi(x_3,y).
\end{equation}
The two equations (\ref{f1}) and (\ref{f2}) are the starting point for the derivation of the QBE for the particle asymmetries. Let us extract from  (\ref{f1}) and (\ref{f2}) the equations of motions for the Green function $G^{<}_{\phi}(x,y)$
\begin{eqnarray}
\left(\stackrel{\rightarrow}{\Box}_x+m^2\right)G^{<}_{\phi}(x,y)&=&
\int\:d^4 x_3\left[ \Sigma^{t}_{\phi}(x,x_3)G^{<}_{\phi}(x_3,y)-\Sigma^{<}_{\phi}(x,x_3)G^{\bar{t}}_{\phi}(x_3,y)\right],\\
G^{<}_{\phi}(x,y)\left(\stackrel{\leftarrow}{\Box}_y+m^2\right)&=&
\int\:d^4 x_3\left[ G^{t}_{\phi}(x,x_3)\Sigma^{<}_{\phi}(x_3,y)-G^{<}_{\phi}(x,x_3)\Sigma^{\bar{t}}_{\phi}(x_3,y)\right].
\end{eqnarray}
If we now substract the  two equations and make the identification  $x=y$, the left-hand side is given by 
\begin{equation}
\left. \partial_\mu^x\left[\left(\partial_x^\mu-\partial_y^\mu\right) G^{<}_{\phi}(x,y)\right]\right|_{x=y}=
-\frac{\partial J_\phi^\mu(X)}{\partial X^\mu}=-\left(\frac{\partial n_\phi}{\partial T}+\stackrel{\rightarrow}{\nabla}_{R}
\cdot\vec{j}_\phi\right),
\end{equation}
and the QBE for the particle density asymmetry is therefore obtained to be
\begin{equation}
\label{con}
\frac{\partial n_\phi(X)}{\partial T}+\stackrel{\rightarrow}{\nabla}_{R}\cdot 
\vec{j}_\phi(X)=-\left. \int\:d^4 x_3\left[\Sigma^{t}_{\phi} G^{<}_{\phi}-\Sigma^{<}_{\phi} G^{\bar{t}}_{\phi}-G^{t}_{\phi} \Sigma^{<}_{\phi}-G^{<}_{\phi} \Sigma^{\bar{t}}_{\phi}\right]\right|_{x=y}.
\end{equation}
In order to examine the ``scattering term'' on the right-hand side of Eq. (\ref{con}), the first step is to restore all the variable arguments. Setting  $t=\vec{r}=0$ in the original notation of $\Sigma_\phi(x,x_3) G_\phi(x_3,y)$ gives  $(X,x_3)(x_3,X)$ for the pair of arguments
\begin{eqnarray}
\label{s}
\frac{\partial n_\phi(X)}{\partial T}+\stackrel{\rightarrow}{\nabla}_{R}\cdot 
\vec{j}_\phi(X)=&-&\int\:d^4 x_3\left[\Sigma^{t}_{\phi}(X,x_3) G^{<}_{\phi}(x_3,X)-\Sigma^{<}_{\phi}(X,x_3) G^{\bar{t}}_{\phi}(x_3,X)\right. \nonumber\\
&+&\left. G^{t}_{\phi} (X,x_3)\Sigma^{<}_{\phi}(x_3,X)-G^{<}_{\phi}(X,x_3) \Sigma^{\bar{t}}_{\phi}(x_3,X)\right].
\end{eqnarray}
The next step is to employ the definitions in (\ref{def1}) to express the time-ordered functions $G^{t}_{\phi}$, $G^{\bar{t}}_{\phi}$, $\Sigma ^t_\phi$, and $\Sigma^{\bar{t}}_{\phi}$ in terms of $G^{<}_{\phi}$, $G^{>}_{\phi}$, 
 $\Sigma^{<}_{\phi}$ and  $G^{>}_{\phi}$.  Then the time integrals are separated into whether
$t_3>T$ or $t_3<T$ and the right-hand side of Eq. (\ref{s}) reads
\begin{eqnarray}
&=&-\int\: d^4 x_3\:\left\{\theta(T-t_3)\left[\Sigma^{>}_{\phi} G^{<}_{\phi}+G^{<}_{\phi}\Sigma^{>}_{\phi}-
\Sigma^{<}_{\phi} G^{>}_{\phi}-G^{>}_{\phi}\Sigma^{<}_{\phi}\right]\right.\nonumber\\
&+&\left. \theta(t_3-T)\left[\Sigma^{<}_{\phi} G^{<}_{\phi}+G^{<}_{\phi}\Sigma^{<}_{\phi}-
\Sigma^{<}_{\phi} G^{<}_{\phi}-G^{<}_{\phi}\Sigma^{<}_{\phi}\right]\right\}.
\end{eqnarray}
The term with $t_3>T$ all cancel, leaving $T>t_3$.  Rearranging these terms gives
\begin{eqnarray}
\label{aaa}
\frac{\partial n_\phi(X)}{\partial T}+\stackrel{\rightarrow}{\nabla}_{R}\cdot 
\vec{j}_\phi(X)&=&-\int\: d^3 r_3\:\int_{-\infty}^{T}\: dt_3\left[\Sigma^{>}_{\phi}(X,x_3) G^{<}_{\phi}(x_3,X)-
G^{>}_{\phi}(X,x_3) \Sigma^{<}_{\phi}(x_3,X)\right.\nonumber\\
&+&\left. G^{<}_{\phi}(X,x_3)\Sigma^{>}_{\phi}(x_3,X)-\Sigma^{<}_{\phi}(X,x_3) G^{>}_{\phi}(x_3,X)\right].
\end{eqnarray}
This equation is the QBE for the particle density asymmetry we were looking for. The right-hand side represents the ``scattering'' term. 
In the particular case in which interactions conserve the number of particles and the latter are neither created nor destroyed, their number asymmetry $n_\phi$ is conserved and should obey the equation of continuity $\partial n_\phi/\partial T+\stackrel{\rightarrow}{\nabla}_{R}\cdot 
\vec{j}_\phi=0$. 
To check that this is indeed the case, one can observe that under the assumption that interactions do not change the number of particles,  most self-energy expressions  can be expressed in the following form 
\begin{equation}
\Sigma^{>}_{\phi}(x,y)=g(x,y) \left[G^{>}_{\phi}(x,y)\right]^m, \:\:\:\:\Sigma^{<}_{\phi}(x,y)=g(x,y) \left[G^{<}_{\phi}(x,y)\right]^m,
\end{equation}
where $g(x,y)=g(y,x)$ and $m$ is a positive integer. This form of the self-energy is found, for instance, for a $\lambda\left|\phi\right|^4$ theory, where $m=3$.  In such a case, the terms in the integrand of the scattering  integral all cancel since 
\begin{equation}
\left\{\left[G^{>}_{\phi}(X,x_3)\right]^3G^{<}_{\phi}(x_3,X)-\left[G^{<}_{\phi}(X,x_3)\right]^3
G^{>}_{\phi}(x_3,X)\right\}\times\left[ g(X,x_3)-g(x_3,X)\right]=0.
\end{equation}
The equation of continuity  is therefore satisfied by the QBE.  In the most interesting cases, however, the particle asymmetries are not conserved in a given environment. This occurs if the 
interactions themselves do not conserve the  particle number asymmetries and  there is some source of CP-violation in the system.  Now, if one follows the spirit of the usual derivation of Fick's law and the diffusion equation, one should perform a simultaneous expansion to first order in the deviations of the distribution function $n_\phi(X)$ from its equilibrium value $n_\phi^0$, in derivatives of  $(n_\phi-n^0_\phi)$ and in the particle number violating interactions.  What is unusual, however,  in Eq. (\ref{aaa}) is the presence of the integral over the time. The physical interpretation of this integral over the past history of the system is straightforward: it leads to the typical ``memory'' effects which are observed in quantum transport theory \cite{dan,henning}. In  the classical kinetic theory the ``scattering term'' does not include any integral over the past history of the system which is equivalent to assume that any collision in the plasma  does not depend upon the previous ones. On the contrary,   
quantum distributions posses strong memory effects and the thermalization rate obtained from quantum transport theory may be substantially longer than the one obtained from classical kinetic theory. This observation is relevant, for instance, when analysing the properties of the quark-gluon plasma \cite{henning1}. We will return to this point in the following. 

The right-hand side of Eq. (\ref{aaa}), through the general form of the self-energy $\Sigma_\phi$,  contains all the informations necessary to describe the temporal evolution of the particle density asymmetries:  particle number
changing reactions and CP-violating source terms,  which will  pop out from the corresponding self-energy $\Sigma_{{\rm CP}}$. Notice that so far we have not made any approximation and the computation is therefore  valid for all
 shapes and sizes of the bubble wall expanding in the thermal bath during a
 first-order electroweak phase transition.
If the interactions of the system do not violate CP,   there will be no CP-violating sources and the final baryon asymmetry produced during supersymmetric baryogenesis will be vanishing. What is noticeable is that we have been able to rigously derive a set of quantum transport equations which  incorporate the CP-violating sources in a self-consistent way.  This is an improvement with respect to recent treatments where the various 
CP-violating currents induced by the wall are first derived and then converted into sources for the diffusion equations.   We will explicitly derive the CP-violating source for the right-handed stop number asymmetry and comment about  its  interpretation as a ``scattering'' term after we have derived  the quantum transport  equations for fermionic particle number asymmetries. 
\begin{flushleft}
{\bf  4. QBE for particle density asymmetry: the fermionic case}
\end{flushleft}

In this Section we will derive the  QBE for the following generic fermionic CP-violating current
\begin{equation}
\langle J_\psi^\mu(x) \rangle \equiv  \langle \bar{\psi}(x)\gamma^\mu \psi(x)\rangle\equiv \left[ n_\psi(x), \vec{J}_\psi(x)\right],
\end{equation}
where $\psi$ indicates  a Dirac fermion  and $\gamma^\mu$ represent the usual Dirac matrices. Again, the zero-component of this current $n_\psi$ represents the number density of particles minus the number density of antiparticles and is therefore the relevant quantity  for the diffusion equations of supersymmetric electroweak baryogenesis.

Our initial goal  is to find a couple of  equation of motions for the interacting fermionic Green function $\widetilde{G}_\psi(x,y)$ when the system is not in equilibrium. Such  equations  may be found  by applying  the operators $\left(i\stackrel{\rightarrow}{\not  \partial}_x -M\right)$
and $\left(i\stackrel{\leftarrow}{\not  \partial}_y +M\right)$  on both sides of  Eqs. (\ref{d1}) and (\ref{d2}), respectively. Here $M$ represents the mass term of the fermion $\psi$. We find
\begin{eqnarray}
\label{c}
\left(i\stackrel{\rightarrow}{\not  \partial}_x -M\right)\widetilde{G}_\psi(x,y)&=&\delta^{(4)}(x,y)\widetilde{I}_4+
\int\:d^4 x_3 \widetilde{\Sigma}_\psi(x,x_3)\widetilde{G}_\psi(x_3,y),\\
\widetilde{G}_\psi(x,y)\left(i\stackrel{\leftarrow}{\not  \partial}_y +M\right)&=& -\delta^{(4)}(x,y)\widetilde{I}_4
-\int\:d^4 x_3 \widetilde{G}_\psi(x,x_3)\widetilde{\Sigma}_\psi(x_3,y).
\end{eqnarray}
We can  now  take the trace over the spinorial indeces of  both sides of the equations, sum up  the two equations above  and finally extract the equation of motion for the Green function $G^{>}_{\psi}$
\begin{eqnarray}
\label{v}
{\rm Tr} \left\{\left[i\stackrel{\rightarrow}{\not  \partial}_x + i\stackrel{\leftarrow}{\not  \partial}_y\right]
G^{>}_{\psi}(x,y)\right\}&=& \int\:d^4 x_3\:{\rm Tr}\left[\Sigma^{>}_\psi(x,x_3)G^t_\psi(x_3,y)-\Sigma^{\bar{t}}_\psi(x,x_3)G^{>}_\psi(x_3,y)\right.\nonumber\\
&-&
\left. G^{>}_\psi(x,x_3)\Sigma^t_\psi(x_3,y)+G^{\bar{t}}_\psi(x,x_3)\Sigma^{>}_\psi(x_3,y)\right].
\end{eqnarray}
If we now make use of the definitions (\ref{dd}),  we can work out the left-hand side of Eq.  (\ref{v})
\begin{eqnarray}
& &\left.{\rm Tr}\left[i\stackrel{\rightarrow}{\not  \partial}_x G^{>}_\psi(T,\vec{R},t,\vec{r})+
G^{>}_\psi(T,\vec{R},t,\vec{r})i\stackrel{\leftarrow}{\not  \partial}_y\right]\right|_{t=\vec{r}=0}\nonumber\\
&=&\left. i\left(  \partial^x_\mu + \partial^y_\mu\right) i \langle
\bar{\psi}\gamma^\mu \psi\rangle\right|_{t=\vec{r}=0}\nonumber\\
&=&-\frac{\partial}{\partial X^\mu} \langle
\bar{\psi}(X)\gamma^\mu \psi(X)\rangle\nonumber\\
&=&-\frac{\partial}{\partial X^\mu} J^\mu_\psi.
\end{eqnarray}
The next step is to employ the definitions in (\ref{def2}) to express the time-ordered functions $G^{t}_{\psi}$, $G^{\bar{t}}_{\psi}$, $\Sigma ^t_\psi$, and $\Sigma^{\bar{t}}_{\psi}$ in terms of $G^{<}_{\psi}$, $G^{>}_{\psi}$, 
 $\Sigma^{<}_{\psi}$ and  $G^{>}_{\psi}$. The computation goes along the same lines of the analysis made in the previous section and we get
\begin{eqnarray}
\label{b}
\frac{\partial n_\psi(X)}{\partial T}+\stackrel{\rightarrow}{\nabla}_{R}\cdot 
\vec{j}_\psi(X)=&-&\int\: d^3 r_3\:\int_{-\infty}^{T}\: dt_3\:{\rm Tr}\left[\Sigma^{>}_{\psi}(X,x_3) G^{<}_{\psi}(x_3,X)-  G^{>}_{\psi}(X,x_3) \Sigma^{<}_{\psi}(x_3,X)     \right.\nonumber\\
&+&\left. G^{<}_{\psi}(X,x_3)\Sigma^{>}_{\psi}(x_3,X)-\Sigma^{<}_{\psi}(X,x_3) G^{>}_{\psi}(x_3,X)\right].
\end{eqnarray}
This is the ``diffusion'' equation describing the temporal evolution of a generic fermionic number asymmetry $n_\psi$. As for the bosonic case, all the informations regarding particle number violating interactions and CP-violating sources  are  stored in the self-energy $\Sigma_\psi$. In the following we will explicitly work out the CP-violating sources for charged and neutral Higgsinos.

\begin{flushleft}
{\bf  5. The  CP-violating source for the right-handed stop number}
\end{flushleft}
 
As we mentioned  in the introduction, a strongly first order electroweak
phase transition
can  be achieved in the presence of a top squark
lighter than the top quark~\cite{r1,r2}.
In order to naturally
suppress its contribution to the parameter $\Delta\rho$ and hence
preserve a good agreement with the precision measurements at LEP,
it should be mainly right handed. This can be achieved if the left
handed stop soft supersymmetry breaking mass $\widetilde{m}_{\widetilde{t}_L}$
is much larger than $M_Z$. Under this assumption, only the right-handed stops contribute to the 
 axial stop charge. 
The right-handed stop current
$J^\mu_{\widetilde{t}_R}$ associated to the right-handed stop $\widetilde{t}_R$
is given by
\begin{equation}
J_{\widetilde{t}_R}^\mu=i\left(\widetilde{t}^*_R\stackrel{\leftrightarrow}
{\partial^\mu}\widetilde{t}_R\right).
\end{equation}
To fix our conventions, let us write the interaction terms among the right-handed stop
$\widetilde{t}_R$, the left-handed stop $\widetilde{t}_L$ and the two neutral Higgses 
$H_{1,2}^0$, which are responsible for  the CP-violating source in the diffusion equation for the right-handed stop number $n_{\widetilde{t}_R}$
\begin{equation}
\label{interaction}
{\cal L}=h_t \widetilde{t}_L\left(A_t H_2^0-\mu^* H_1^0\right)\widetilde{t}_R^*+ {\rm h.c.}.
\end{equation}
Here the soft trilinear term $A_t$ and the supersymmeric mass term $\mu$ are meant to be complex parameters so that ${\rm Im}(A_t\mu)$ is nonvanishing.  Even though in this paper   we will restrict ourselves to the computation of  the CP-violating source in the diffusion equation of  the particle asymmetry $n_{\widetilde{t}_R}$, it is clear that the self-energy of the right-handed stop contains the informations about all the  
other interactions which are responsible for changing $n_{\widetilde{t}_R}$ in the plasma. A typical example is provided by the interaction among the right-handed stop, the left-handed top $t_L$ and the Higgsino $\widetilde{H}_2^0$.  Eq. (\ref{aaa}) is the QBE describing  the right-handed stop number asymmetry. Solving this equation  represents an Herculean task  since it is     integral  and nonlinear.  This happens because the self-energy functions $\Sigma^{>}$ and $\Sigma^{<}$ are also functions of the full nonequilibrium Green functions
of other degrees of freedom of the system.  We can make some progress, though. 
Since we know that   there is no CP-violating source in the diffusion equation of  $n_{\widetilde{t}_R}$ in absence of any Higgs configuration describing the bubble wall profile, we first perform a  ``Higgs insertion expansion'' around the symmetric phase $\langle H_i^0(x)\rangle=v_i(x)=0$ $(i=1,2)$. 
At the lowest level of perturbation, the interactions (\ref{interaction}) induce a contribution to 
the  self-energy of the form
\begin{equation}
\label{q}
\Sigma_{{\rm CP}}^{>}(x,y)=g_{{\rm CP}}(x,y)G^{0,>}_{\widetilde{t}_L}(x,y), \:\:\:\:
\Sigma_{{\rm CP}}^{<}(x,y)=g_{{\rm CP}}(x,y)G^{0,<}_{\widetilde{t}_L}(x,y),
\end{equation}
where $G^{0,>}_{\widetilde{t}_L}$ and $G^{0,<}_{\widetilde{t}_L}$ are now  the Green functions for the left-handed stop computed in the {\it unbroken} phase and
\begin{equation}
\label{u}
g_{{\rm CP}}(x,y)=h_t^2 \left[A_t^* v_2(x)-\mu v_1(x)\right] \left[A_t v_2(y)-\mu^* v_1(y)\right].
\end{equation}
If we now insert the expressions (\ref{q}) and  (\ref{u}) into the diffusion equation
 (\ref{aaa}), we get
\begin{equation}
\label{p}
 \frac{\partial n_{\widetilde{t}_R}}{\partial T}+\stackrel{\rightarrow}{\nabla}_{R}\cdot 
\vec{j}_{\widetilde{t}_R}={\cal S}_{\widetilde{t}_R}+\cdots.
\end{equation}
where
\begin{eqnarray}
{\cal S}_{\widetilde{t}_R}&=&- 2 i \int\: d^3 r_3\:\int_{-\infty}^{T}\: dt_3\left[g_{{\rm CP}}(X,x_3)-g_{{\rm CP}}(x_3,X)\right]\nonumber\\
&\times& {\rm Im}\left[G^{0,>}_{\widetilde{t}_L}(X,x_3) G^{0,<}_{\widetilde{t}_R}(x_3,X)\right]+\cdots\nonumber\\
&=& 4\:h_t^2 \int\: d^3 r_3\:\int_{-\infty}^{T}\: dt_3 \:{\rm Im}\left( A_t\mu\right)\left[v_2(X)v_1(x_3)-
v_2(x_3) v_1(X)\right]\nonumber\\
&\times& {\rm Im}\left[G^{0,>}_{\widetilde{t}_L}(X,x_3) G^{0,<}_{\widetilde{t}_R}(x_3,X)\right]+\cdots.
\end{eqnarray}
where the dots represent the other terms describing the particle number violating interactions. 
${\cal S}_{\widetilde{t}_R}$ is   the CP-violating source for the right-handed stop number asymmetry. Notice that it vanishes if the relative phase of $A_t\mu$ is zero and if the ratio ${\rm tan}\beta(x)\equiv v_2(x)/v_1(x)$ is a constant in the interior of the bubble wall.  The corresponding diagram is given in Fig. 3 where the thick dashed line stands for the fact that one has to compute the imaginary part of the diagram.  The interpretation of the CP-violating source as a ``scattering'' term is therefore   straightforward: the CP-violating source is built up when the  right-handed stops pass across the wall, they first scatter off the wall and are transformed into left-handed stops; the latter subsequently suffer another scattering off the wall and are converted again into right-handed stops. If  CP-violation is taking place in both interactions, a nonvanishing 
CP-violating source ${\cal S}_{\widetilde{t}_R}$ pops out from thermal bath. 

In order to deal with analytic expressions, we can work out
the thick wall limit and simplify the expressions obtained above
by performing a derivative expansion
\begin{equation}
\label{expansion}
v_i(x_3)= \sum_{n=0}^{\infty}\frac{1}{n!}\; \frac{\partial^n}
{\partial (X^\mu)^n} v_i(X)\left(x^\mu_3-X^\mu\right)^n .
\end{equation}
The term
with no
derivatives vanishes in the expansion (\ref{expansion}),
$v_2(X)v_1(X)-v_1(X)v_2(X)= 0$, which means that the static
term in the derivative expansion  does not contribute
to the source  ${\cal S}_{\widetilde{t}_R}$.
For a smooth Higgs profile, the derivatives with
respect to the time coordinate and $n>1$ are associated with higher
powers of $v_{\omega}/L_{\omega}$,  where $v_{\omega}$ and $L_{\omega}$ are the velocity and the width of the bubble wall, respectively.  Since the typical time scale of the processes giving rise to the source is given by the thermalization time of the stops $1/\Gamma_{\widetilde{t}}$, 
the approximation is good for values of
$L_{\omega}\Gamma_{\stop}/v_{\omega} \gg 1$.
In other words, this expansion is valid only when the mean free path of the stops in the plasma 
 is smaller than the scale of variation of the Higgs
background determined by the wall thickness, $L_{\omega}$,
and the wall velocity $v_{\omega}$. A detailed 
computation of the   thermalization rate of the  right-handed stop from the imaginary
 part of the two-point Green function has been recently performed in \cite{therm} by making use of improved propagators and including
 resummation of hard thermal loops\footnote{The left-handed stop is usually considered to be much heavier than $T$ and its decay width corresponds to the one in the present vacuum.}. The thermalization rate has been  computed exactly at the
 one-loop level in the high temperature approximation as a function of the plasma right-handed stop mass 
 $m_{\widetilde{t}_R}(T)$ and an  estimate for the magnitude of the
 two-loop contributions which dominate the rate for small $m_{\widetilde{t}_R}(T)$ was also given. 
If  $m_{\widetilde{t}_R}(T)\gsim  T$, the thermalization  is dictated by the one-loop thermal decay rate which can be larger than $T$ \cite{therm}\footnote{For smaller values of   $m_{\widetilde{t}_R}(T)$, when the 
 thermalization is dominated by two-loop effects ({\it i.e.} scattering),   $\Gamma_{\widetilde{t}_R}$ may be as large as $10^{-3} T$ \cite{therm}.}. 
 With such value, our derivative expansion is
perfectly justified since the wall thickness can span the range
$(10-100)/T$. 

The  term corresponding to  $n=1$  in the expansion (\ref{expansion})  gives a contribution to the source proportional to the function 
\begin{equation}
v_1(X)\partial_X^\mu v_2(X)- v_2(X)\partial_X^\mu v_1(X)
\equiv v^2(X) \partial_X^\mu\beta(X),
\end{equation}
which  should vanish smoothly for values of $X$ outside the
bubble wall. Here  we have denoted $v^2\equiv v_1^2+ v_2^2$. 
Since the variation of the Higgs fields is due to  the
expansion of the bubble wall through the thermal bath,
the source ${\cal S}_{\widetilde{t}_R}$
will be linear in $v_{\omega}$.
This result explicitly shows that we need out of equilibrium
conditions to generate the source and that we
have
to call for the CTP formalism to deal with time-dependent
phenomena.
To work out exactly  ${\cal S}_{\widetilde{t}_R}$  one should
know the exact form of the Green functions
which, in ultimate analysis,
are provided by solving the complete set of Quantum  Boltzmann equations.
However, any departure from thermal equilibrium distribution
functions
is caused at a given point by the passage of the wall and, therefore,
is  ${\cal O}(v_{\omega})$.  Since the source  is
already linear in $v_{\omega}$,
working with thermal {\it equilibrium} Green 
functions
amounts to ignoring terms of higher order in
$v_{\omega}$. This is 
 as accurate as the bubble wall is moving slowly in
the plasma. 

The generic finite temperature, real-time propagator in {\it  equilibrium}
$G^{t}_\phi(\k,t_x-t_y)$ can be written in terms of the spectral
function $\rho_\phi(\k,k_0)$ \cite{ww}
\begin{equation}
G^{0,t}_\phi(\k,t_x-t_y)=\int_{-\infty}^{+\infty}\:
\frac{d k^0}{2\pi}\:{\rm e}^{-i k^0(t_x-t_y)}\:\rho_\phi(\k,k^0)\:\left\{
\left[1+n_\phi(k^0)\right]\theta(t_x-t_y)+n_\phi(k^0)\theta(t_x-t_y)\right\},
\label{rho}
\end{equation}
where $n_\phi(k^0)$ represents the Bose-Einstein distribution function.

To account for
interactions with the surrounding particles of the thermal bath,
particles must be
substituted by quasiparticles,  dressed propagators are to be adopted
(the use of
the full corrected propagators should be done with some care to avoid an
overcounting of diagrams \cite{par}) and 
 self-energy
corrections at one- or two-loops to
the propagator modify the dispersion relations by 
introducing a finite width $\Gamma_\phi(k)$. 
In the limit of small decay width,  the spectral function is expressed by
\begin{equation}
\rho_\phi(\k,k^0)=i\:\left[\frac{1}{(k^0+i\varepsilon+ i\Gamma_\phi)^2-\omega_\phi^2(k)}-
\frac{1}{(k^0-i\varepsilon-i\Gamma_\phi)^2-\omega_\phi^2(k)}\right],
\end{equation}
where $\omega_\phi^2(k)=\k^2 +m_\phi^2(T)$ and $m_\phi(T)$ is the thermal mass. 
Performing the integration over $k^0$ 
one gets \cite{ww}
\begin{eqnarray}
\label{a}
G_\phi^{0,>}(\k,t_x-t_y)&=&-\frac{1}{2\:\omega_\phi}\left\{
\left[1+n(\omega_\phi-i\Gamma_\phi)\right]\:{\rm
e}^{-i(\omega_\phi-i\Gamma_\phi)(t_x-t_y)}+n(\omega_\phi+i\Gamma_\phi)\:
{\rm
e}^{-i(\omega_\phi+i\Gamma_\phi)(t_x-t_y)}\right\},\nonumber\\
G_\phi^{0,<}(\k,t_x-t_y)&=&G_\phi^{0,>}(\k,t_y-t_x).
\end{eqnarray}
Since the Green functions   depend only  upon the absolute value of the three-momentum,  the contribution  to the source ${\cal S}_{\widetilde{t}_R}$ from the  $n=1$ term  in the derivate expansion  (\ref{expansion})   vanishes when we select  the space coordinates  $(\mu=1,2,3)$. Indeed, in such  a case the source is proportional to 
\begin{equation}
\int \frac{d^3{\bf k}}{(2\pi)^3}\: \delta^{(3)}({\bf k})\:\frac{\partial}{\partial {\bf k}} G^{0,>}_{\widetilde{t}_L}({\bf k}, t_X-t_3)\equiv 0.
\end{equation}
We are therefore left with the expression corresponding to Fig.3
\begin{equation}
\label{source1}
{\cal S}_{\widetilde{t}_R}(X)=h_t^2\:{\rm Im}\left( A_t\mu\right) v^2(X)\dot{\beta}(X)\:{\cal I}_{\widetilde{t}_R},
\end{equation}
where $\dot{\beta}(X)\equiv d\beta(X)/dt_X$, 
\begin{eqnarray}
{\cal I}_{\widetilde{t}_R} &=& \int_0^\infty dk \frac{k^2}{2 \pi^2  \;
\omega_{\widetilde{t}_L} \; \omega_{\widetilde{t}_R}}   \nonumber\\
&\times& \left[ \left(1 + 2 {\rm Re}(n_{\widetilde{t}_L}) \right)
I(\omega_{\widetilde{t}_R},\Gamma_{\widetilde{t}_R},\omega_{\widetilde{t}_L},\Gamma_{\widetilde{t}_L})
+
\left(1 + 2 {\rm Re}(n_{\widetilde{t}_R}) \right)
I(\omega_{\widetilde{t}_L},\Gamma_{\widetilde{t}_L},\omega_{\widetilde{t}_R},\Gamma_{\widetilde{t}_R})\right.  \nonumber\\
&-&\left. 
2 \left( {\rm Im}(n_{\widetilde{t}_R}) +
{\rm Im}(n_{\widetilde{t}_L}) \right) G(\omega_{\widetilde{t}_R},\Gamma_{\widetilde{t}_R},
\omega_{\widetilde{t}_L},\Gamma_{\widetilde{t}_L}) \right]
\end{eqnarray}
and 
$n_{\widetilde{t}_R(L)} = 1/\left[\exp\left(\omega_{\widetilde{t}_R(L)}/T + i \Gamma_{\widetilde{t}_R(L)}/T
\right)- 1 \right]$. 
The functions $I$ and $G$ are given by 
\begin{eqnarray}
\label{v}
I(a,b,c,d) &=& \frac{1}{2}\frac{1}{\left[(a+c)^2 + (b+d)^2 \right]}
\sin\left[ 2{\rm arctan}\frac{a+c}{b+d}\right]\nonumber\\
&+&\frac{1}{2}\frac{1}{\left[(a-c)^2 + (b+d)^2 \right]}
\sin\left[ 2{\rm arctan}\frac{a-c}{b+d}\right],\nonumber\\
G(a,b,c,d)=&-&\frac{1}{2}\frac{1}{\left[(a+c)^2 + (b+d)^2 \right]}
\cos\left[ 2{\rm arctan}\frac{a+c}{b+d}\right]\nonumber\\
&-&\frac{1}{2}\frac{1}{\left[(a-c)^2 + (b+d)^2 \right]}
\cos\left[ 2{\rm arctan}\frac{a-c}{b+d}\right].
\end{eqnarray}
Notice that the function $G(\omega_{\widetilde{t}_R},\Gamma_{\widetilde{t}_R},
\omega_{\widetilde{t}_L},\Gamma_{\widetilde{t}_L})$ has a peak for $\omega_{\widetilde{t}_R}\sim\omega_{\widetilde{t}_L}$.  
This resonant behaviour \cite{noi}  is associated to the fact that 
the Higgs background  is
carrying a very low momentum (of order of the inverse of the bubble wall
width $L_\omega$) and to the 
possibility of absorption or emission of Higgs quanta by the
propagating supersymmetric particles. The resonance  can only take place when  the left-handed stop and the right-handed stop  do not differ too much in mass.
By using the Uncertainty Principle, it is easy to understand that the
width of this resonance is expected
to be proportional to the thermalization rate  of the particles giving rise to
the baryon asymmetry.  
Within the MSSM, however, it is assumed that  $\widetilde{m}_ {\widetilde{t}_L}\gg T$ and the resonance  can only happen for
momenta larger than  $\widetilde{m}_ {\widetilde{t}_L}$. Such configurations are
exponentially suppressed and do not give any relevant
contribution to the CP-violating source. Nonertheless, if the  electroweak phase transition is enhanced by the presence of some
new degrees of freedom beyond the ones contained in the MSSM, {\it
e.g.} some extra standard model gauge singlets, the resonance effects in the stop sector might be relevant.  What is relevant here is that the source may be  dominated by particles with long wavelengths in direction perpendicular to the wall for which the classical approximation breaks down.

\begin{flushleft}
{\bf  6. The  CP-violating source for the Higgsino number}
\end{flushleft}
 
The Higgs fermion  current associated with  neutral
and charged Higgsinos can be written
as
\be
\label{corhiggs}
J^{\mu}_{\widetilde{H}}=\overline{\widetilde{H}}\gamma^\mu \widetilde{H}
\ee
where $\widetilde{H}$ is the Dirac spinor
\be
\label{Dirac}
\widetilde{H}=\left(
\begin{array}{c}
\widetilde{H}_2 \\
\overline{\widetilde{H}}_1
\end{array}
\right)
\ee
and $\widetilde{H_2}=\widetilde{H}_2^0$ ($\widetilde{H}_2^+$),
$\widetilde{H_1}=\widetilde{H}_1^0$ ($\widetilde{H}_1^-$) for
neutral (charged) Higgsinos. The interactions among the charginos and the charged Higgsinos which are responsible for the CP-violating source in the diffusion equation for the Higgs fermion number read
\begin{equation}
{\cal L}=-g_2\left\{\overline{\widetilde{H}}\left[v_1(x)P_L+{\rm e}^{i\theta_\mu} v_2(x) P_R\right]\widetilde{W}\right\}+{\rm h.c.},
\end{equation}
where $\theta_\mu$ is the phase of the $\mu$-parameter and  $P_{L,R}$ are the chirality projector operators. Analogously, the interactions among the Bino, the $\widetilde{W}_3$-ino and the neutral Higgsinos are
\begin{equation}
 {\cal L}=-\frac{1}{2}\left\{\overline{\widetilde{H}^0}\left[v_1(x)P_L+{\rm e}^{i\theta_\mu} v_2(x) P_R\right]\left(g_2\widetilde{W}_3-g_1\widetilde{B}\right)\right\}+{\rm h.c.}.
\end{equation}
To compute the source for the Higgs fermion number ${\cal S}_{\widetilde{H}}$ 
 we again  perform a  ``Higgs insertion expansion'' around the symmetric phase. At the lowest level of perturbation, the interactions of the charged Higgsino induce 
a contribution to the self-energy of the form (and analogously for the other component 
$\Sigma_{{\rm CP}}^{>}$)
\begin{equation}
\label{qf}
\Sigma_{{\rm CP}}^{<}(x,y)=g^L_{{\rm CP}}(x,y)P_L G^{0,<}_{\widetilde{W}}(x,y) P_L+
g^R_{{\rm CP}}(x,y)P_R G^{0,<}_{\widetilde{W}}(x,y) P_R,
\end{equation}
where 
\begin{eqnarray}
\label{qs}
g^L_{{\rm CP}}(x,y)&=&g_2^2 v_1(x) v_2(y){\rm e}^{-i\theta_\mu},\nonumber\\
g^R_{{\rm CP}}(x,y)&=&g_2^2 v_1(y) v_2(x){\rm e}^{i\theta_\mu}.
\end{eqnarray}
Similar formulae hold for the neutral Higgsinos. 

Analogously to the case of right-handed stops, the dispersion
relations of charginos and neutralinos are changed by high
temperature
corrections~\cite{weldon}. Even though fermionic dispersion
relations
are highly nontrivial,  especially when dealing with Majorana fermions \cite{majorana}, relatively simple expressions
for the equilibrium fermionic spectral functions may be given in the limit
in which the damping rate is smaller than the typical self-energy
of the fermionic excitation ~\cite{henning}. For instance, the spectral
function of  the charged Higgsinos   may be written as
\begin{eqnarray}
\rho_{\widetilde{H}}({\bf k},k^0)& = & i
\left(\not k + m_{\widetilde{H}}\right) \\
&&
\left[\frac{1}{(k^0+i\varepsilon+ i
\Gamma_{\widetilde{H}})^2-\omega_{\widetilde{H}}^2(k)}-
\frac{1}{(k^0-i\varepsilon-i\Gamma_{\widetilde{H}})^2
-\omega_{\widetilde{H}}^2(k)}\right],\nonumber
\label{rofermion}
\end{eqnarray}
where $\omega_{\widetilde{H}}^2(k)={\bf k}^2 +
m_{\widetilde{H}}^2(T)$ and $m_{\widetilde{H}}^2(T)$
is the Higgsino effective plasma squared mass in the thermal bath
which
may be well approximated by its value in the present vacuum,
$m_{\widetilde{H}}^2(T)\simeq |\mu|^2$. Similarly, $|\mu|$
should
be replaced by $M_2$ for $\rho_{\widetilde{W}}({\bf k},k^0)$,
and by $M_1$ for $\rho_{\widetilde{B}}({\bf k},k^0)$.
Inserting the expressions (\ref{qf}) and (\ref{qs}) into the diffusion equation (\ref{b}) ,  we can perform
a Higgs insertion expansion of the CP-violating source.  The computation goes along the same lines of the calculation done in the previous section and it is easy to show that 
the CP-violating source 
\begin{eqnarray}
{\cal S}_{\widetilde{H}}&=&- \int\: d^3 r_3\:\int_{-\infty}^{T}\: dt_3\:{\rm Tr}\left[\Sigma_{{\rm CP}}^{>}(X,x_3) G^{<}_{\widetilde{H}}(x_3,X)-
G^{>}_{\widetilde{H}}(X,x_3) \Sigma_{{\rm CP}}^{<}(x_3,X)\right.\nonumber\\
&+&\left. G^{<}_{\widetilde{H}}(X,x_3)\Sigma_{{\rm CP}}^{>}(x_3,X)-\Sigma_{{\rm CP}}^{<}(X,x_3) G^{>}_{\widetilde{H}}(x_3,X)\right],
\end{eqnarray}
containes in the integrand the following function
\begin{equation}
g^L_{{\rm CP}}(X,x_3)+g^R_{{\rm CP}}(X,x_3)-
g^L_{{\rm CP}}(x_3,X)-g^R_{{\rm CP}}(x_3,X)=2i\sin\theta_\mu\left[v_2(X)v_1(x_3)-v_1(X)v_2(x_3)\right],
\end{equation}
which vanishes if  ${\rm Im}(\mu)=0$ and if the $\tan\beta(x)$ is a constant  along the Higgs profile. Performing the "Higgs derivative expansion", we finally get 
\begin{equation}
\label{source2}
{\cal S}_{\widetilde{H}}(X) =
{\rm Im}(\mu)\: \left[ v^2(X)\dot{\beta}(X) \right]
\left[ 3 M_2 \; g_2^2 \; {\cal I}^{\widetilde{W}}_{\widetilde{H}}
 +       M_1 \; g_1^2 \; {\cal I}^{\widetilde{B}}_{\widetilde{H}}
\right],
\end{equation}
where
\begin{eqnarray}
{\cal I}^{\widetilde{W}}_{\widetilde{H}} & = & \int_0^\infty dk
\frac{k^2}
{2 \pi^2 
\omega_{\widetilde{H}} \omega_{\widetilde{W}}} \nonumber\\
&\left[ \phantom{\frac{1}{2^2}} \right.&
 \left(1 - 2 {\rm Re}(n_{\widetilde{W}}) \right)
I(\omega_{\widetilde{H}},\Gamma_{\widetilde{H}},
\omega_{\widetilde{W}},\Gamma_{\widetilde{W}})+
\left(1 - 2 {\rm Re}(n_{\widetilde{H}}) \right)
I(\omega_{\widetilde{W}},
\Gamma_{\widetilde{W}},\omega_{\widetilde{H}},
\Gamma_{\widetilde{H}}) \nonumber\\
&+&
2 \left( {\rm Im}(n_{\widetilde{H}}) +
{\rm Im}(n_{\widetilde{W}}) \right)
G(\omega_{\widetilde{H}},
\Gamma_{\widetilde{H}},
\omega_{\widetilde{W}},\Gamma_{\widetilde{W}})
\left.\phantom{\frac{1}{2^2}} \right]
\nonumber\\
\end{eqnarray}
and $\omega^2_{\widetilde{H}(\widetilde{W})}=k^2+ |\mu|^2
(M_2^2)$ while $n_{\widetilde{H}(\widetilde{W})} =
1/\left[\exp\left(\omega_{\widetilde{H}(\widetilde{W})}/T
+ i \Gamma_{\widetilde{H}(\widetilde{W})}/T \right)
+ 1 \right]$.
The exact computation of the  damping rate of charged and neutral
Higgsinos  will be given elsewhere \cite{inprep}. 
The Bino contribution may be obtained from the above
expressions by replacing $M_2$ by $M_1$. As  for ${\cal S}_{\widetilde{t}_R}$, the CP-violating source for the Higgs fermion number is enhanced  if   $M_{2}, M_{1}\sim \mu$ and  low momentum particles are transmitted over the distance $L_\omega$. 
This means that    the classical approximation is not  entirely adequate to
describe the quantum interference nature of $CP$-violation and only a quantum approach is suitable for the computation of the building up of the CP-violating sources.  

{\bf  7.  Outlook}
Let us now look back and comment about the various aspects of our findings. 

\begin{flushleft}
{ \it  --Comparison to previous  work--}
\end{flushleft}

One of the merits of the CTP formalism is to guide us towards a rigorous and self-consistent definition of the CP-violating sources  {\it within}  the quantum Boltzmann equations.  On the contrary, previous treatements \cite{nelson,noi}  are characterized by the  following common feature:  
 CP-violating currents  were first derived and then  coverted    into sources for the diffusion equations.  This procedure is (at least theoretically) not   self-consistent.  More specifically, CP-violating sources ${\cal S}$ associated to a generic charge density $j^0$ were constructed from 
the current $j^\mu$  by the definition ${\cal S}=\partial_0 j^0$ \cite{nelson,noi}.  A rigorous computation of the CP-violating currents for the right-handed stop and higgsino local densities  was  performed in \cite{noi} by means of the CTP formalism. Since   currents  were proportional to $\dot{\beta}$ in the tick bubble wall limit, {\it i.e.}  proportional to the first time derivative of the the Higgs profile,  sources turned out to be   proportional to the second time derivative of the Higgs profile \cite{noi}.  Our results, however, indicate that the sources in the quantum diffusion equations are proportional to the first time derivative of the Higgs configuration.   A comparison between the sources ${\cal S}_{\widetilde{t}_R}$, see Eq.   (\ref{source1}),  and ${\cal S}_{\widetilde{H}}$, see Eq.  (\ref{source2}),  obtained in the present work and the 
currents $j^0$ given in Eqs.   (14) and (21)  of ref. \cite{noi} indicate that  they may be related
as
\begin{equation}
\label{z}
{\cal S}(T)\sim \frac{j^0(T)}{\tau}
\end{equation}
and it may be interpreted as the time derivative of the current density accumulated at the time
$T$ after  the wall has deposited at a given specific point the current density $j^0$ each interval $\tau$
\begin{equation}
{\cal S}(T)\sim \partial_0 \int^T dt \:\frac{j^0(t)}{\tau}.
\end{equation}
Here 
 $\tau=\Gamma^{-1}$ is the thermalization time of the right-handed stops and higgsinos, respectively. The integral over time is peculiar of the quantum approach and it induces memory effects. 
This tells us that the source obtained self-consistently in the present work differs from the one adopted in \cite{noi} by a  factor $\sim  L_\omega\Gamma/v_\omega$ (in the rest frame of the advancing bubble wall). Since 
$L_\omega\Gamma/v_\omega\simgt 1$  for  the Higgs derivative expansion to hold, this result is important as far as the numerical estimate of the final baryon number is concerned.  We notice that the definition of the source given in \cite{nelson} is very similar to (\ref{z}) even though it was not motivated by first principles and it did not incorporate  self-consistently 
 the decoherence effects which have a crucial impact on the generation of the CP-violating observables.  

{\it  --Memory effects--}

Baryogenesis is fueled  when transport properties 
allow the CP-violating  
charges to efficiently diffuse in front of the advancing bubble wall where
anomalous electroweak baryon violating processes are not suppressed.
However, the CP-violating processes of quantum interference,  which build up  
CP-violating sources,  must  act in opposition to  the incoherent nature of  
plasma physics  responsible for the loss of quantistic interference.  
If the particles  involved in the process of baryon number generation 
thermalize rapidly, CP-violating   
sources 
loose their coherence and are diminished. The  CTP formalism properly describes the quantum nature of CP-violation and tells us that   
CP-violating sources evaluated at some time $T$  are always  proportional to an integral over the past history of the system.  Therefore, it is fair to  argue that these memory effects  lead to ``relaxation'' times  for the CP-violating sources which   are   typically longer than the ones dictated by the thermalization rates of the particles in the  thermal bath.  In fact, this observation
is valid for all the processes described by the ``scattering'' term in the right-handed side of the quantum diffusion equations. 
The slowdown of the relaxation processes  may help to keep the system out of equilibrium for longer times and therefore enhance the final baryon asymmetry. There are  two more  reasons why one should expect  quantum relaxation times to be  longer than the ones predicted by the classical approach. First, the decay of the Green's functions  as functions of the difference of the time arguments: an exponential decay is found in thermal equilibrium when one ignore the frequency dependence of self-energies in the spectral functions,   {\it e.g.}  $\left| G^{>}({\bf k},t,t^\prime)\right|\sim \left| G^{>}({\bf k})\right| \times {\rm exp}\left[-\Gamma({\bf k},\omega)|t-t^\prime|\right]$. The decay of the Green's functions restrict the range of the time integration for the scattering term, reduces the integrals and, therefore, the change  of the local particle number densities as a function of time.  The second effect is the rather different oscillatory behaviour of the functions $G^{>}$ and $G^{<}$ for a given momentum, as functions of the time argument difference. 

{\it  --Resonance effects--}

In the limit of thick bubble walls,   the CP-violating sources are characterized by resonance effects \cite{noi} when the particles involved in the construction of the source are degenerate in mass.  The resonance is manifest in the function $G$ defined in (\ref{v}). The interpretation of the resonance is rather straightforward if we think  in terms of  scatterings of the quasiparticles off the advancing low momentum bubble wall configuration.  A similar effect has been found in ref. \cite{ck} where the system was  studied  in the classical limit. These classical treatments somehow obscure the origin of the CP-violating effects as resulting from quantum interference and the origin of the resonance is less transparent to us.  However, these methods
should provide reasonable approximations to our formulae to those particles whose wavelength
is short compared to $v_\omega/\Gamma$. On the other hand,  formulae should not agree for small $\Gamma$ because our source is dominated by particles with long wavelength. In this regime,   the classical approximation breaks down since it  requires that  the mean free path should  be larger than  the Compton
 wavelength of the underlying particle.  This is relevant because  quasiparticles with long wavelengths  give a significant contribution to CP-violating sources.

\vskip1cm
{\large\bf Acknowledgements}
\vskip 0.2cm
The author would like to thank M. Carena, J. Cline,  M. Quiros and C.E.M. Wagner for useful discussions and in particular R. Kolb whose never-ending skepticism about he idea of electroweak baryogenesis spurred, is spurring and will always spur his efforts. 
He  would also like to thank the Theoretical Astrophysics group at  Fermilab and the Particle Theory group at Johns Hopkins, Baltimore,  where part of this work was done, for their warm hospitality.

\newpage
{\large\bf Figure Captions}
\vskip 1cm

{\bf Fig. 1 }: The first Dyson equation for the Green function matrix $\widetilde{G}$. Here thick  and thin solid lines represent the Green functions for the fully interacting system and for the free theory, respectively. 

\vskip 0.3cm

{\bf Fig. 2 }: The second Dyson equation for the Green function matrix $\widetilde{G}$. The meaning of  thick and thin solid lines is as in Fig. 2. 

\vskip 0.3cm

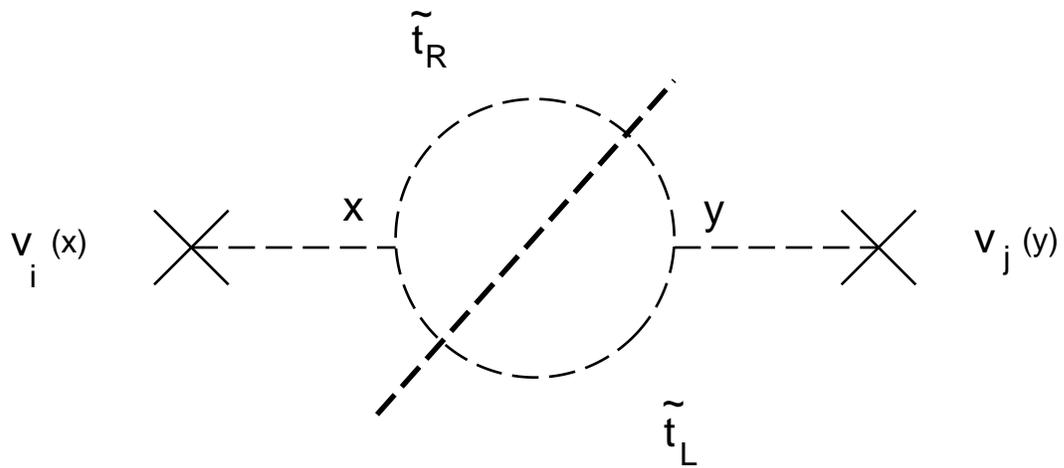
{\bf Fig. 3  }:  The Feynman diagram representing the CP-violating source for the right-handed stop number. The indeces $i,j$ run from 1 to 2 and the solid dashed line means that the imaginary part of  the diagram should be considered. 

\vskip 0.3cm

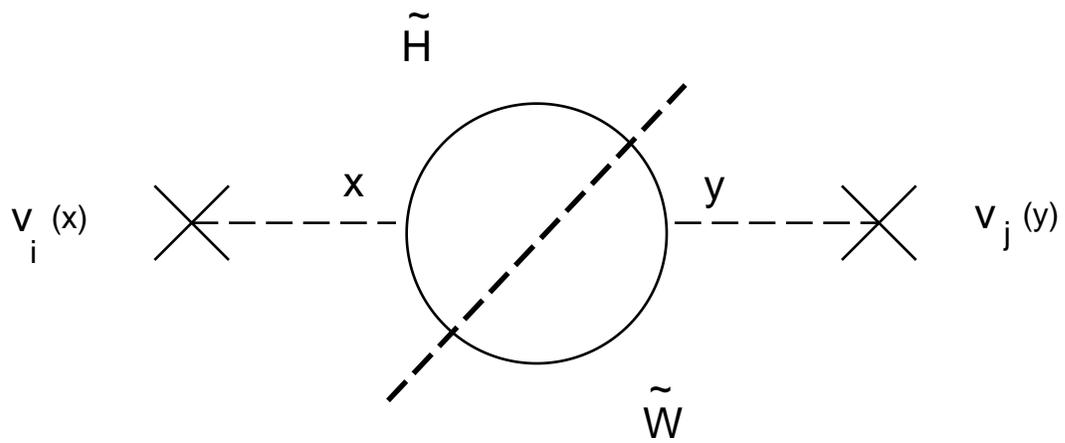
{\bf Fig. 4  }:  The Feynman diagram representing the CP-violating source for the Higgsino number. The indeces $i,j$ run from 1 to 2 and the solid dashed line means that the imaginary part of  the diagram should be considered.

\def\NPB#1#2#3{Nucl. Phys. {\bf B#1}, #3 (19#2)}
\def\PLB#1#2#3{Phys. Lett. {\bf B#1}, #3 (19#2) }
\def\PLBold#1#2#3{Phys. Lett. {\bf#1B} (19#2) #3}
\def\PRD#1#2#3{Phys. Rev. {\bf D#1}, #3 (19#2) }
\def\PRL#1#2#3{Phys. Rev. Lett. {\bf#1} (19#2) #3}
\def\PRT#1#2#3{Phys. Rep. {\bf#1} (19#2) #3}
\def\ARAA#1#2#3{Ann. Rev. Astron. Astrophys. {\bf#1} (19#2) #3}
\def\ARNP#1#2#3{Ann. Rev. Nucl. Part. Sci. {\bf#1} (19#2) #3}
\def\MPL#1#2#3{Mod. Phys. Lett. {\bf #1} (19#2) #3}
\def\ZPC#1#2#3{Zeit. f\"ur Physik {\bf C#1} (19#2) #3}
\def\APJ#1#2#3{Ap. J. {\bf #1} (19#2) #3}
\def\AP#1#2#3{{Ann. Phys. } {\bf #1} (19#2) #3}
\def\RMP#1#2#3{{Rev. Mod. Phys. } {\bf #1} (19#2) #3}
\def\CMP#1#2#3{{Comm. Math. Phys. } {\bf #1} (19#2) #3}


\end{document}